\documentclass[iop,apj,tighten,times]{emulateapj}

\usepackage{apjfonts} 
\usepackage{amsmath,amstext}
\usepackage{times}
\usepackage{graphicx}
\usepackage{latexsym,amssymb}
\usepackage{amsfonts,exscale}

\DeclareGraphicsExtensions{.pdf,.png,.jpg}

\shorttitle{Violent quenching}
\shortauthors{J.~E.~Geach et al.}

\newcommand{\co}{CO\,{\it J}(2$\rightarrow$1)}

\begin{document}

\title{Violent quenching: molecular gas blown to 1000\,\lowercase{km\,s$^{-1}$} during a major merger}

\author{J.~E.~Geach$^1$,
C.~Tremonti$^2$,
A.~M.~Diamond-Stanic$^3$, 
P.~H.~Sell$^{4,5}$,
A.~A.~Kepley$^6$,\\
A.~L.~Coil$^7$,
G.~Rudnick$^8$, 
R.~C.~Hickox$^9$,
J.~Moustakas$^{10}$ \&
Yujin~Yang$^{11}$\vspace{9pt}}

\affil{$^1$Centre for Astrophysics Research, School of Physics, Astronomy \& Mathematics, University of
Hertfordshire, Hatfield, AL10 9AB, UK. j.geach@herts.ac.uk}
\affil{$^2$Department of Astronomy, University of Wisconsin-Madison, 475 N. Charter St., Madison, WI, 53706, USA}
\affil{$^3$Department of Physics and Astronomy, Bates College, 44 Campus Ave, Carnegie Science Hall, Lewiston, ME 04240, USA}
\affil{$^4$Foundation for Research and Technology-Hellas, 71110 Heraklion, Crete, Greece}
\affil{$^5$Physics Department \& Institute of Theoretical \& Computational Physics, University of Crete, 71003 Heraklion, Crete, Greece}
\affil{$^6$National Radio Astronomy Observatory, 520 Edgemont Road, Charlottesville, VA 22903, USA}
\affil{$^7$Center for Astrophysics and Space Sciences,
  Department of Physics, University of California, 9500 Gilman Dr., La
  Jolla, CA 92093, USA}
\affil{$^8$Department of Physics and Astronomy, The University of Kansas, Malott Room 1082, 1251 Wescoe Hall Drive, Lawrence, KS, 66045, USA}
\affil{$^9$Department of Physics and Astronomy, Dartmouth College, 6127 Wilder Laboratory, Hanover, NH 03755, USA}
\affil{$^{10}$Department of Physics and Astronomy, Siena College, 515 Loudon Road, Loudonville, NY 12211, USA}
\affil{$^{11}$ Korea Astronomy and Space Science Institute,
776 Daedeokdae-ro, Yuseong-gu, Daejeon, Korea 34055}

\begin{abstract}
We present Atacama Large Millimeter/submillimeter Array 
observations of a massive ($M_\star\approx10^{11}M_\odot$) compact ($r_{\rm e, UV}\approx 100$\,pc) merger remnant at $z=0.66$ that is driving a 1000\,km\,s$^{-1}$ outflow of cool gas, with no observational trace of an active galactic nucleus (AGN). We resolve molecular gas on scales of approximately 1--2\,kpc, and our main finding is the discovery of a wing of blueshifted \co{} emission
out to $-$1000\,km\,s$^{-1}$ relative to the stars. We argue that this is the molecular component of a multiphase outflow, expelled from the central starburst within the past 5\,Myr through stellar feedback, although we cannot rule out previous AGN activity as a launching mechanism. If the latter is true, then this is an example of a relic multiphase AGN outflow. We estimate a molecular mass outflow rate of approximately 300\,$M_\odot$\,yr$^{-1}$, or about one third of the 10\,Myr--averaged star formation rate. This system epitomizes the multiphase `blowout' episode following a dissipational major merger -- a process that has violently quenched central star formation and supermassive black hole growth.
\end{abstract}

\section{Introduction}

When accurate measurements of the mean baryon density of the Universe were made ($\Omega_{\rm b}\approx0.04$, e.g.\ Spergel et al.\ 2003), it was quickly realised that the standard model of galaxy formation of the day (e.g.\ Rees \& Ostriker\ 1977; White \& Rees 1978; White \& Frenk 1991; Somerville \& Primack\ 1999) needed urgent revision. In particular, truncation, or at least dramatic modification, of the star formation histories of massive galaxies is required in models of galaxy formation (Benson et al.\ 2003; Granato et al.\ 2004; De Lucia et al.\ 2005) to reproduce the demographics of the galaxy population at $z=0$ (e.g.\ Bower et al.\ 2006; Croton et al.\ 2006).

When star formation is curtailed 
more rapidly than what would be expected from normal consumption of the gas reservoir, we refer to this as `quenching'. At its most extreme end, feedback from energy and momentum input to the interstellar medium might rapidly (within a few 10s Myr) shut down star formation on whole galaxy
scales (Silk \& Rees\ 1998). The most efficient means of achieving this is by rapidly destroying, or removing, molecular gas (Feruglio et al.\ 2010; Sturm et al.\ 2011; Veilleux et al.\ 2013; Bolatto et al.\ 2013; Geach et al.\ 2014; Biernacki \& Teyssier 2018; Cicone et al.\ 2014, 2018). We refer to this process as `violent quenching'. This is relevant not only to the star formation history of galaxies but also their chemical evolution, as it propels  metal-enriched gas through galaxies and into the circumgalactic medium (Veilleux, Cecil \& Bland-Hawthorn\ 2013; Tumlinson, Peeples \& Werk 2017; Baron et al.\ 2018).

In galaxy formation lore (e.g.\ Silk \& Rees\ 1998; Di Matteo et al.\ 2005), feedback from active galactic nuclei (AGN) is the standard route to quench massive galaxies, with quasar mode feedback capable of violent quenching through blowout (Hopkins \& Elvis\ 2010). However, we have been studying a rare sample of massive ($M_\star\approx 10^{11}M_{\odot}$) compact galaxies at $z\approx0.6$ driving ultrafast outflows of cool gas that appear to be driven by pure stellar feedback (Diamond-Stanic et al.\ 2012).

Our sample was selected from the Sloan Digital Sky Survey (SDSS, York et al.\ 2000) to have spectral features indicative of early-stage quenching.  Subsequent follow-up observations revealed highly blueshifted ($\Delta V\approx -1000$\,km\,s$^{-1}$) Mg\,{\sc ii} $\lambda\lambda$2796,~2804\AA\ interstellar absorption lines indicative of unusually fast ionized gas outflows (Tremonti et al.\ 2007). \emph{Hubble Space Telescope} ({\it HST}) imaging reveals that the galaxies have incredibly compact optical morphologies with effective radii of $r_{\rm e}\approx100$\,pc, and tidal features suggestive of late-stage major mergers (Sell et al.\ 2014). Their {\it WISE} 22$\mu$m fluxes indicate high star formation rates (SFRs) and imply extreme SFR surface densities $\Sigma_{\rm SFR}>500$\,$M_\odot$\,yr$^{-1}$\,kpc$^{-2}$ (Diamond-Stanic et al.\ 2012), well in excess of what is required to launch such extreme winds (e.g.\ Murray et al.\ 2005). The majority of these systems present no clear evidence of an energetically-dominant AGN (Sell et al. 2014), suggesting stellar feedback could be driving the outflows. 

Geach et al.\ (2013 \& 2014) detected \co{} emission in two of the sample, revealing an implied global star formation efficiency close to the upper limit of Eddington-limited models of star formation (Murray et al.\ 2005; Thompson et al.\ 2015), and the presence of a high velocity \co{} wing, indicating a molecular outflow. In this Letter, we present new, more sensitive, \co{{} observations of another galaxy in the sample with the  Atacama Large Millimeter/submillimeter Array (ALMA). We are able to resolve the \co{} on scales down to 0.25$''$, allowing us to map the molecular gas on similar scales as the starlight. Throughout we assume a {\it Planck} 2015 cosmology, $h=H_0/100\,{\rm km\,s^{-1}\,Mpc^{-1}}=0.677$, $\Omega_{\rm m}=0.307$, $\Omega_\Lambda=0.693$ (Planck Collaboration\ 2015).

\section{Observations and data reduction}

SDSS\,J1341$-$0321 ($z=0.66$) at $\alpha=13^{\rm h}$41$^{\rm m}$36.80$^{\rm s}$,
 $\delta=-$03$^\circ$21$'$25.3$''$ (J2000) was observed in ALMA Cycle 4 (2016.1.01072.S) and in Cycle 5 (2017.1.01318.S). The observations are summarized in Table\ 1.

\begin{table}[h]
\caption{ALMA observations of SDSS\,J1341$-$0321}
\vspace{-12pt}
\begin{center}
\begin{tabular}{lcc}
\hline
Campaign & Cycle 4 & Cycle 5 \cr
\hline
Dates of observation & Mar 8 2017 & Dec 30 2017 \cr
& & Jan 6,\,7 2018 \cr 
Number of 12\,m antennas & 41 & 43--46 \cr
Baseline separation & 15--331\,m & 15--2517\,m\cr
System temperature $\langle T_{\rm sys} \rangle$ & 85.3\,K & 71.5\,K\cr
Mean pwv  & 4.8\,mm & 2.2\,mm \cr
Average elevation & 61$^\circ$ & 62$^\circ$\cr
Time on source & 123\,min & 292\,min\cr
Atmospheric calibrators & J1337$-$1257 & J1337-1257 \cr 
& J1335$-$0511 & \cr
& Ganymede\cr 
& Titan   \cr
Bandpass calibrators & 	J1337$-$1257 & 	J1337$-$1257 \cr
Flux calibrators & 	Ganymede & J1337$-$1257 \cr
& Titan & \cr
Phase calibrators & 	J1335$-$0511 & 	J1332$-$0509  \cr
Pointing calibrators & 	J1337$-$1257  &  J1337$-$1257\cr 
& J1742$-$1517 \cr
& J1332$-$0509 & \cr
\hline
\end{tabular}
\end{center}
\end{table}

We concatenate the calibrated Cycle 4 and 5 measurement sets in {\sc
  casa} (version 5.1.0) and image them using the {\it clean}
command. To explore the morphology of the source we image the
visibilities using natural and Briggs weighting, with robust
parameter $R = 0$ for the latter. First we generate dirty
maps to identify the \co{} emission and to evaluate the channel
sensitivity. We then clean to a stopping threshold of 1$\sigma$, using
a circular mask of radius 5$''$ centered on the source (amply covering
the emission). We employ multi-scale cleaning with four components corresponding to a delta function and Gaussians with widths of 1, 2 and 5$\times$ the synthesized natural beam, which has a full width at half maximum of $(0.42''\times0.32'')$ with position angle of 88\,degrees.

\section{Analysis \& Results}

\subsection{Optical and near-infrared imaging}

In Figure~1 we present {\it HST} imaging of SDSS\,J1341$-$0321, revealing a compact central core with two faint tidal arms spanning several arcseconds. The target
was observed with the Wide Field Camera 3 (WFC3)
Ultraviolet Imaging Spectrograph (UVIS) channel using the F475W and
F814W filters and the infrared (IR) channel
 using the F160W filter.  All images were
registered to one another with {\it tweakreg} to $\delta_\theta\approx0.01''$ and then
boresight-aligned to an SDSS (Albareti et al. 2017) overlapping field
to the approximate astrometric precision of SDSS ($\delta_\theta\approx0.2''$).

Like most other galaxies in our sample, SDSS\,J1341$-$0321 is extremely compact. When quantifying the compactness with {\sc galfit} (version 3; Peng et al. 2010), we focus on the F475W and F814W bands, which have the best spatial resolution ($\textnormal{FWHM}\approx0.07''$), and we find $r_{\rm e,UV} \simeq 100, 110$~pc based on single-component S\'ersic models with $n=4$. Further details on image processing and morphological analysis can be found in Sell et al.\ (2014).

\begin{figure}
\centerline{\includegraphics[width=\linewidth]{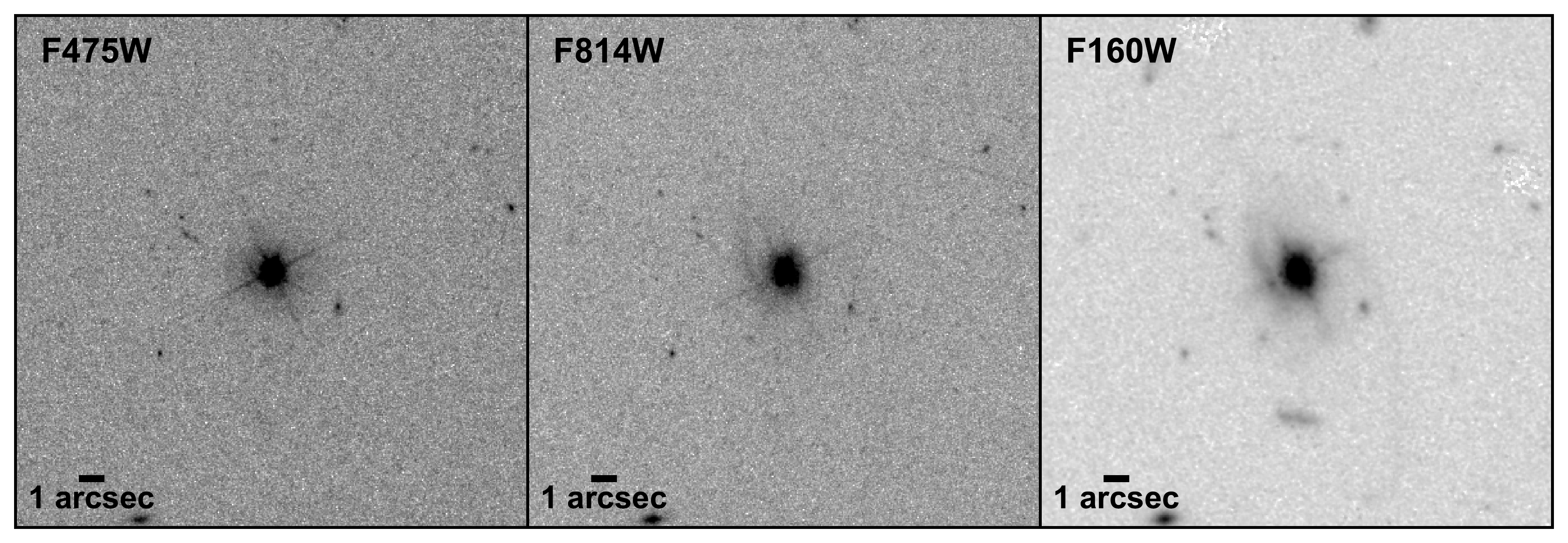}}
\centerline{\includegraphics[width=\linewidth]{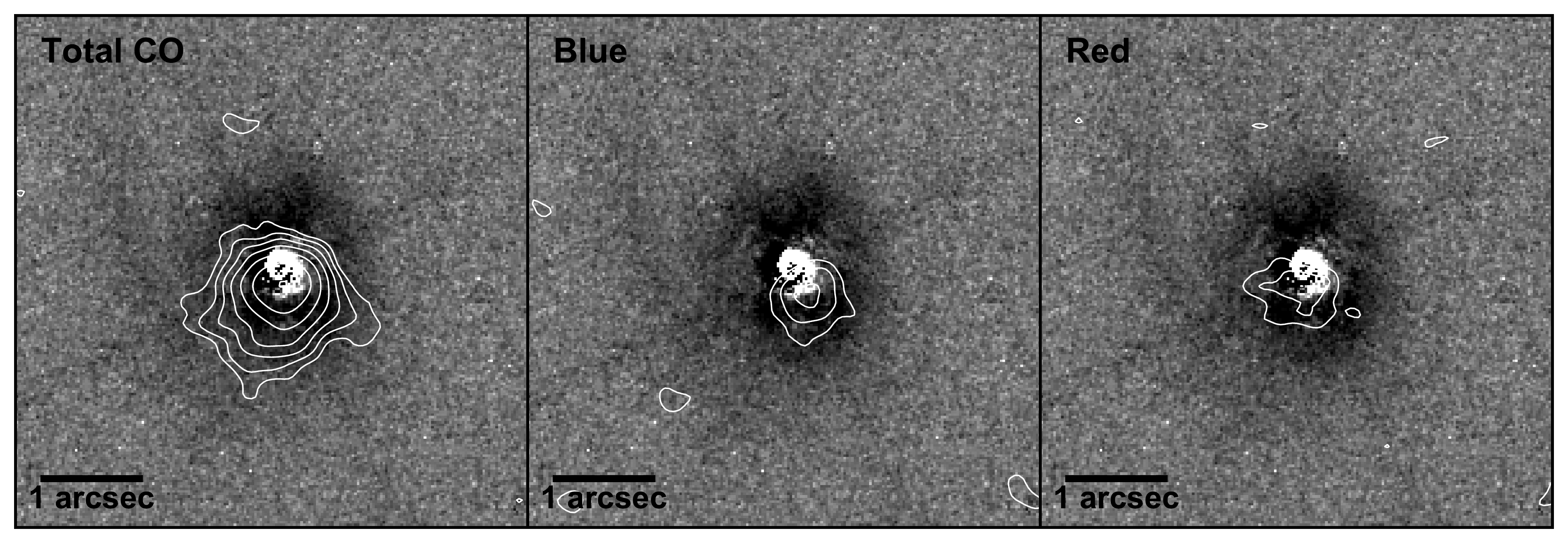}}
\caption{(top) {\it HST} WFC3 UVIS (F475W, F814W) and IR (F160W) images of
  SDSS\,J1341$-$0321 spanning 20$''$. Two faint tidal tails are
  visible and extend over several arcseconds ($1''\approx7$\,kpc), but the majority of the stellar emission is unresolved. (bottom) 5$''$ zoom-ins showing the S\'ersic model-subtracted F814W image (section 3.2) as greyscale and \co{} contours, averaged over (left) $|\Delta V|<1000$\,km\,s$^{-1}$, (middle) $-1000 < \Delta V < -500$\,km\,s$^{-1}$, (right) $500 < \Delta V < 1000$\,km\,s$^{-1}$. \co{} contours start at 3$\sigma$ and are logarithmically spaced at 0.2\,dex multiples of $\sigma$. All images are orientated north up, east left. Note that the absolute astrometric precision of the {\it HST} images (boresight-aligned to SDSS) is $\delta_\theta\approx0.2''$.}
\end{figure}

\begin{figure*}[t]
\centerline{\includegraphics[width=\linewidth,angle=0]{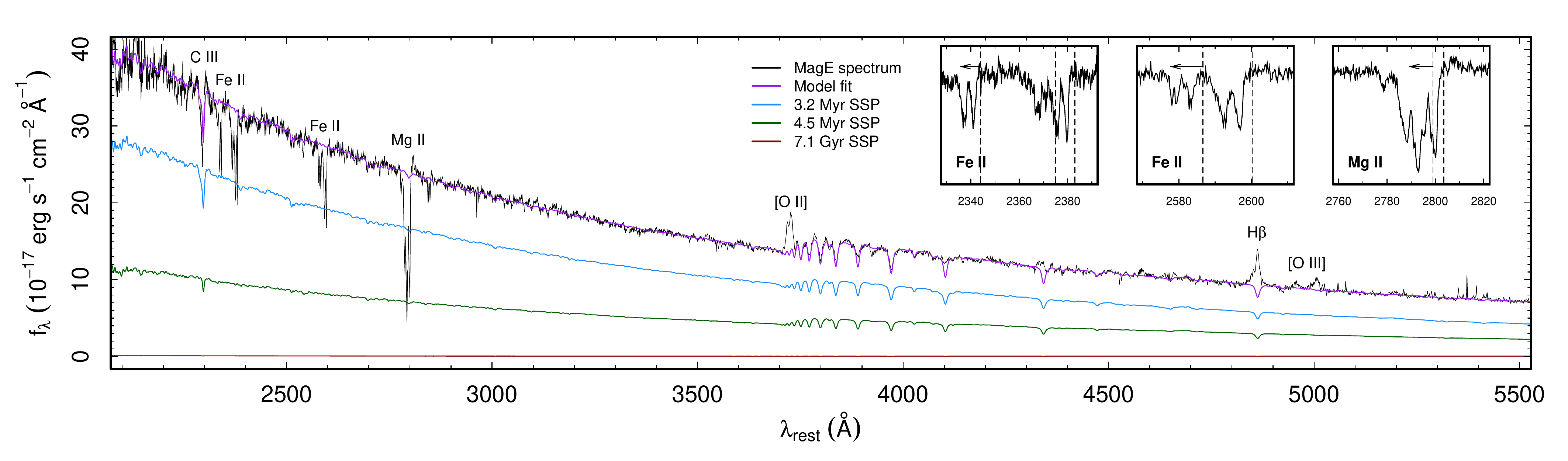}}
\caption{Magellan Echellette (MagE) optical spectrum of SDSS\,J1341$-$0321 in the rest-frame. We show the best fitting model from our SSP modelling (Section 3.2) illustrating that the stellar continuum emission is dominated by very young ($t<5$\,Myr) populations. The redshift $z=0.6611$ is determined from the stellar absorption lines. The inset panels show zoom-ins of the model continuum normalized spectrum around the Fe~{\sc ii}\,$\lambda$2344, 2374, 2383, 2587 \& 2600\AA{} and Mg~{\sc ii\,}$\lambda\lambda$2796,\,2804\AA{} ISM absorption lines. Vertical dashed lines indicate the rest-frame wavelengths of the lines, and the arrows indicate a blueshift vector of $-$1000\,km\,s$^{-1}$. Note that the 7.1\,Gyr component is so weak relative to the young populations it appears almost featureless in this plot.} 
\end{figure*}

\subsection{Optical spectroscopy and stellar population modelling}

Figure\ 2 presents the rest-frame UV--optical spectrum of
SDSS\,J1341$-$0321 obtained with the Magellan Echellette (MagE) spectrograph (Marshall et al.\ 2008) on the Magellan Clay telescope with a 1$''$ slit and 2 hours of integration time.  The data were reduced and calibrated using the {\it MASE} pipeline (Bochanski et al.\ 2009). The spectrum has a resolution $R\sim4100$ over a bandpass of 3300--9400\AA{} and a signal-to-noise of $\sim$45 per resolution element near the galaxy's Mg\,{\sc ii} $\lambda\lambda$2796,2804\AA{} absorption lines. The spectrum displays a very blue UV--optical continuum with weak stellar Balmer absorption lines, suggesting a stellar population dominated by O and B stars. The Mg~{\sc ii} and Fe~{\sc ii} ($\lambda$2344, 2374, 2383, 2587 and 2600\AA{}) ISM absorption lines are highly blueshifted relative to the stars (insets). The presence of two sets of absorption profiles for each species implies two distinct velocity components at roughly $-500$\,km\,s$^{-1}$ and $-1000$\,km\,s$^{-1}$, indicating two potential outflow events.

Could the central unresolved emission, blue, nearly-featureless continuum, and moderately broad H$\beta$ emission line be due to the presence of an unobscured (Type~{\sc i}) AGN? The continuum is very well fit by stellar population models, as shown in Figure~2, and the presence of a strong C~{\sc iii}~$\lambda$2298\AA{} P~Cygni absorption feature indicates a dominant population of Wolf--Rayet stars (Leitherer et al.\ 2014). The H$\beta$ emission line is noticeably broadened ($\sigma \sim600$~km~s$^{-1}$), however it is not as broad as the H$\beta$ lines found in Type~{\sc i} AGNs ($>$$1000$~km~s$^{-1}$; Osterbrock \& Mathews 1986; Sulentic et al.\ 2000). We hypothesize that some of the H$\beta$ emission arises in the outflow itself (see Figure~4). The [O~{\sc ii}] $\lambda\lambda3726,3729$\AA{} line is also dominated by a broad component that could trace the outflow; this line is typically narrow in Type~{\sc i} AGN.  

Could SDSS\,J1341$-$0321 host an obscured (Type~{\sc ii}) AGN?  We combine the MagE spectrum with Keck/NIRSPEC data covering the [N~{\sc ii}]\,$\lambda$6584\AA{} and H$\alpha$ lines (George et al., in prep.), decomposing the lines into broad ($\sim$600\,km\,s$^{-1}$) and narrow ($\sim$200\,km\,s$^{-1}$) components, and place them on the Baldwin, Phillips \& Terlevich (1981) `BPT' AGN diganostic diagram. Both the narrow and broad components of the emission lines fall in the `composite' region (Kewley et. al 2006), likely due to the contribution of shocks associated with the merger and outflow (Rich et al.\ 2014; Ho et al.\ 2014).
Notably, the narrow component lies very close to the star formation divider line when the latter is adjusted for redshift (Kewley et al.\ 2013). Finally, the galaxy's {\it WISE} (Wright et al.\ 2010) {\it W1}$-${\it W2} color is 0.258$\pm$0.005, well below the AGN selection threshold of Stern et al.\ 2012 
({\it W1}$-${\it W2}$>$0.8) and the more conservative threshold recommended by  Blecha et al.\ 2018 ({\it W1}$-${\it W2}$>$0.5).  We conclude that SDSS\,J1341$-$0321 is unlikely (currently) to host an AGN.

We fit the spectrum with a combination of simple stellar population (SSP) models and a Calzetti (2000) reddening law.  We employed the Flexible Stellar Population Synthesis code (Conroy, Gunn \& White 2009; Conroy \& Gunn 2010) to generate SSPs  with Padova\ 2008 isochrones, a Salpeter (1955) initial mass function, and a new theoretical stellar library `C3K' (Conroy et al., in prep) with a resolution of $R\sim10,000$. We utilize solar metallicity SSP templates with 43 ages spanning 1~Myr--8.9~Gyr.  We perform the fit with the Penalized Pixel-Fitting (pPXF) software (Cappellari \& Emsellem 2004; Cappellari 2017). The best fit model is dominated by very young stellar populations, with approximately 87\% of the continuum emission at 5500\AA{} contributed by populations less than 5\,Myr old. The 10\,Myr-averaged SFR inferred from the SSP modelling is
$M_\star\approx900$\,$M_\odot$\,yr$^{-1}$, after converting to a Chabrier IMF (Chabrier et al.\ 2003).   

Given the youth of the stellar population ($t<5$~Myr), the nebular emission lines appear surprisingly weak. The H$\beta$ equivalent width is 10\AA{}, a factor of $\sim$4 lower than SDSS galaxies with comparable colors and redshift. This is not an artifact of differential attenuation, for the stars and the gas have comparable reddening. One possibility is that the compact starburst has an unusually high ionization parameter. This would result in Lyman continuum photons penetrating deeply into the surrounding nebula and potentially being absorbed by dust before ionizing hydrogen atoms. Another possibility is that the outflow blew away much of the natal nebula on a short timescale, allowing ionizing photons to escape from the center of the galaxy. Notably, the Mg~{\sc ii} and Fe~{\sc ii} lines show no absorption near zero velocity, suggesting that the bulk of the ionized gas in the galaxy is outflowing.

\begin{figure*}
\centerline{\includegraphics[width=\linewidth,angle=0]{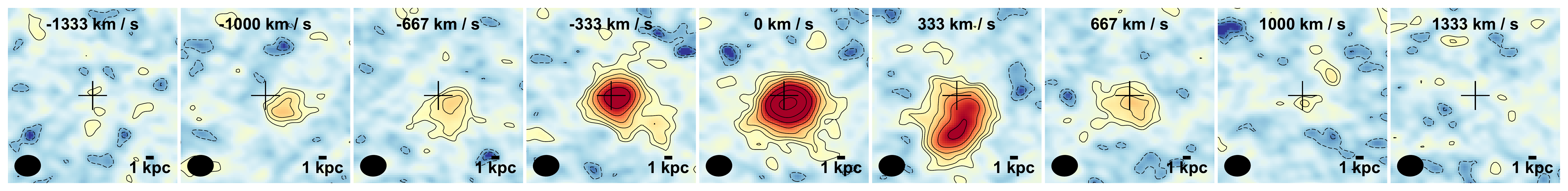}}
\centerline{\includegraphics[width=\linewidth,angle=0]{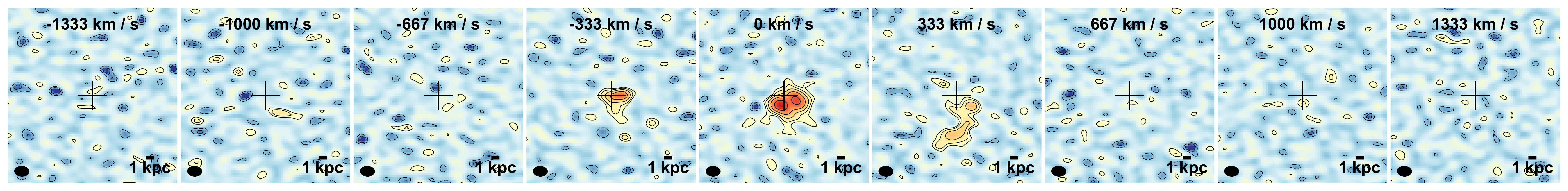}}
\caption{\co{} maps of SDSS\,J1341$-$0321 averaged over 333\,km\,s$^{-1}$ channels spanning
  $\Delta V=\pm1500$\,km\,s$^{-1}$ (labels give channel centers)
  relative to the systemic redshift. The top row shows the naturally weighted images and the bottom row shows the Briggs ($R=0$) weighted images. Contours start at 2$\sigma$ and are logarithmically spaced at 0.2\,dex multiples of $\sigma$. Dashed contours are negative equivalents. The crosshair indicates the peak of the stellar emission in the {\it HST}/F814W imaging (Figure\ 1). Black ellipses show the {\sc fwhm} of the synthesized beam.}
\end{figure*}

 \subsection{\co{} emission}

Figure\ 1 presents the line-integrated \co{} emission in comparison to the {\it HST} imaging and Figure\ 3 shows the \co{} velocity-averaged maps in
333\,km\,s$^{-1}$-wide channels, spanning $|\Delta V| \lesssim 1500$\,km\,s$^{-1}$ around
the systemic redshift of the galaxy. Figure\ 4 shows the total \co{} spectrum. The integrated
line flux is $S_{\rm CO}\Delta V = 3.4\pm0.1$\,Jy\,km\,s$^{-1}$,
corresponding to a luminosity of $L'_{\rm CO}=(2.0\pm
0.1)\times10^{10}$\,K\,km\,s$^{-1}$\,pc$^2$. 

There is strong \co{} emission out to  $\Delta V \approx \pm300$\,km\,s$^{-1}$ relative to the stars (including an inner spiral or tidal arm visible on the receding side), as well a significant wing to $\Delta V \approx-1000$\,km\,s$^{-1}$. In Figure\ 4 we show on the same velocity scale the continuum-subtracted Fe~{\sc ii}
and Mg~{\sc ii} absorption lines and the H$\beta$ and [O~{\sc ii}] nebular emission lines. The nebular and absorption lines extend to similar blueshifted velocities as seen in
\co{} emission, implying they could all be tracing the same outflowing gas. 

The \co{} morphology is complex; in the naturally weighted image
the peak of the emission within a few 100\,km\,s$^{-1}$ of the
systemic is (a) not coincident with the peak of the starlight, and (b)
elongated roughly S--W. A Briggs weighting of the visibilities allows us to examine the central \co{} morphology in more detail by tailoring the `robust' parameter $R$ to
produce a smaller synthesized beam, albeit at the expense of
sensitivity, such that we only detect \co{} emission within $\Delta |V| \lesssim
350$\,km\,s$^{-1}$. With $R=0$ we can image the
\co{} emission at a resolution of $(0.25''\times 0.18'')$. Interestingly, at this resolution the
emission at zero velocity has a double lobed morphology,
straddling the bright peak of the stellar emission. This suggests that
the bulk of the molecular gas close to zero velocity is not coincident with the site of very recent intense star formation. We postulate we are seeing the blue starburst though a relatively low obscuring column -- a result of ISM blowout.

\section{Interpretation}

SDSS\,J1341$-$0321 is clearly a late-stage merger, with obvious tidal features in
the extended stellar emission, and so it is reasonable to conclude
that some of the dynamical structure in the \co{} emission is
associated with dense interstellar material dispersed by tidal
forces. Indeed, as evident in the channel maps (Figure\ 3), there is a prominent
\co{} feature located at $\Delta V \approx 300$\,km\,s$^{-1}$,
describing an arc of molecular emission that is roughly co-spatial
with the inner part of a spiral or tidal arm visible in the {\it HST}
imaging. However, the \co{} emission at $-1000$\,km\,s$^{-1}$ cannot
plausibly be associated with purely tidal debris resulting from the
merger; we argue it is more likely to have been propelled to high
velocity as part of a multiphase outflow driven by the starburst. 

\subsection{Violent quenching}

We estimate the molecular mass outflow rate assuming a time-averaged thin shell approximation (Rupke et al.\ 2005), where \co{} is tracing outflowing H$_2$ with
total mass $M_{\rm H_2,out}$ in a shell of radius $R_{\rm out}$, traveling
at $v_{\rm out}$: $\dot{M}_{\rm H_2,out} = v_{\rm out}M_{\rm H_2,out}/R_{\rm out}$. To estimate $M_{\rm H_2,out}$ we only consider emission between $-500$\,km\,s$^{-1}$ and $-1500$\,km\,s$^{-1}$  and assume the optically thin case appropriate for a turbulent outflow $M_{\rm H_2} = 0.34L'_{\rm CO}$ (Bolatto et al.\ 2013). Obviously the choice of $\alpha_{\rm CO}$ will have a systematic
effect on the derived mass outflow rate: increasing it will increase $\dot{M }_{\rm H_2}$. We further assume that the
\co{} emission is thermalized, which is also a conservative assumption
since a correction for sub-thermal excitation will also drive up the mass estimate. 

Another key uncertainty is the geometry of the outflowing gas. At
$\Delta V\approx -1000$\,km\,s$^{-1}$ the peak of the \co{} emission is offset $(0.28\pm 0.02)''$ from the peak of the \co{} emission measured at zero velocity, corresponding to a projected distance of $(2.0\pm 0.1)$\,kpc, similar to the distance covered by a parcel of gas travelling at 500--1000\,km\,s$^{-1}$ for 3\,Myr, and so we take it as reasonable estimate of $R_{\rm out}$. We fully acknowledge that poor constraints on the wind geometry are a major systematic uncertainty in the calculation. The final consideration is the choice of $v_{\rm out}$. The observed \co{} velocities are subject to projection effects. One approach is to take the maximum velocity in the wing as the deprojected velocity of the outflow (e.g.\ Maiolino et al.\ 2012), however this could potentially dramatically overestimate the outflow rate. We take a conservative approach and do not deproject the velocities, and instead integrate over the spectrum such that
\begin{equation}
\dot{M}_{\rm H_2,out} \propto \int^{\rm -500\,km\,s^{-1}}_{\rm -1500\,km\,s^{-1}}vS_{\nu}{\rm d}v
\end{equation}
This gives $\dot{M}_{\rm H_2,out}=310\pm 70$\,$M_\odot$\,yr$^{-1}$, implying a mass loading factor of $\eta\sim 1/3$ given the 10\,Myr-averaged SFR.

The corresponding kinetic power of the outflow (integrating similarly over the wing) is $P_{\rm k} =
v^2\dot{M}_{\rm H_2,out}/2$, yielding $P_{\rm k}=(0.8\pm
0.3)\times10^{44}$\,erg\,s$^{-1}$, or approximately 1\% of the
bolometric luminosity (cf.\,Feruglio et al.\ 2010), with $\log_{10}(L_{\rm IR}/{\rm
  erg\,s^{-1}})=46.0$ based on the integrated $\lambda_{\rm rest}=8$--$1000$\,$\mu$m Chary \&
Elbaz (2001) template fits to the {\it WISE} 12$\mu$m and 22$\mu$m
flux (Diamond-Stanic et al.\ 2012). The momentum flux in the
molecular gas is $\dot{p} = v\dot{M}_{\rm H_2}$, with
$\dot{p}=(1.8\pm0.5)\times10^{36}$\,dyne. The total momentum input from the recent starburst, $L/c$, is approximately $8\times10^{35}$\,dyne, which seems insufficient to drive the outflow, assuming a momentum-conserving wind. However, allowing for multiple scatterings, additional momentum contributions from supernovae, uncertainties on the wind geometry (particularly the radius) and $\alpha_{\rm CO}$ could bring the values to parity. 
Since there is likely to be additional mass and momentum in the warm and hot phases of the outflow, the momentum budget remains tight.  It is possible that the outflow could have been launched by a past AGN that now leaves no observational trace.

Regardless of the launching mechanism, the high mass outflow rate in molecular gas, young unobscured central starburst remnant and lack of present AGN indicates that the core of this galaxy has been violently quenched, curtailing both stellar mass and supermassive black hole growth.   

\begin{figure}
\centerline{\includegraphics[width=0.9\linewidth,angle=0]{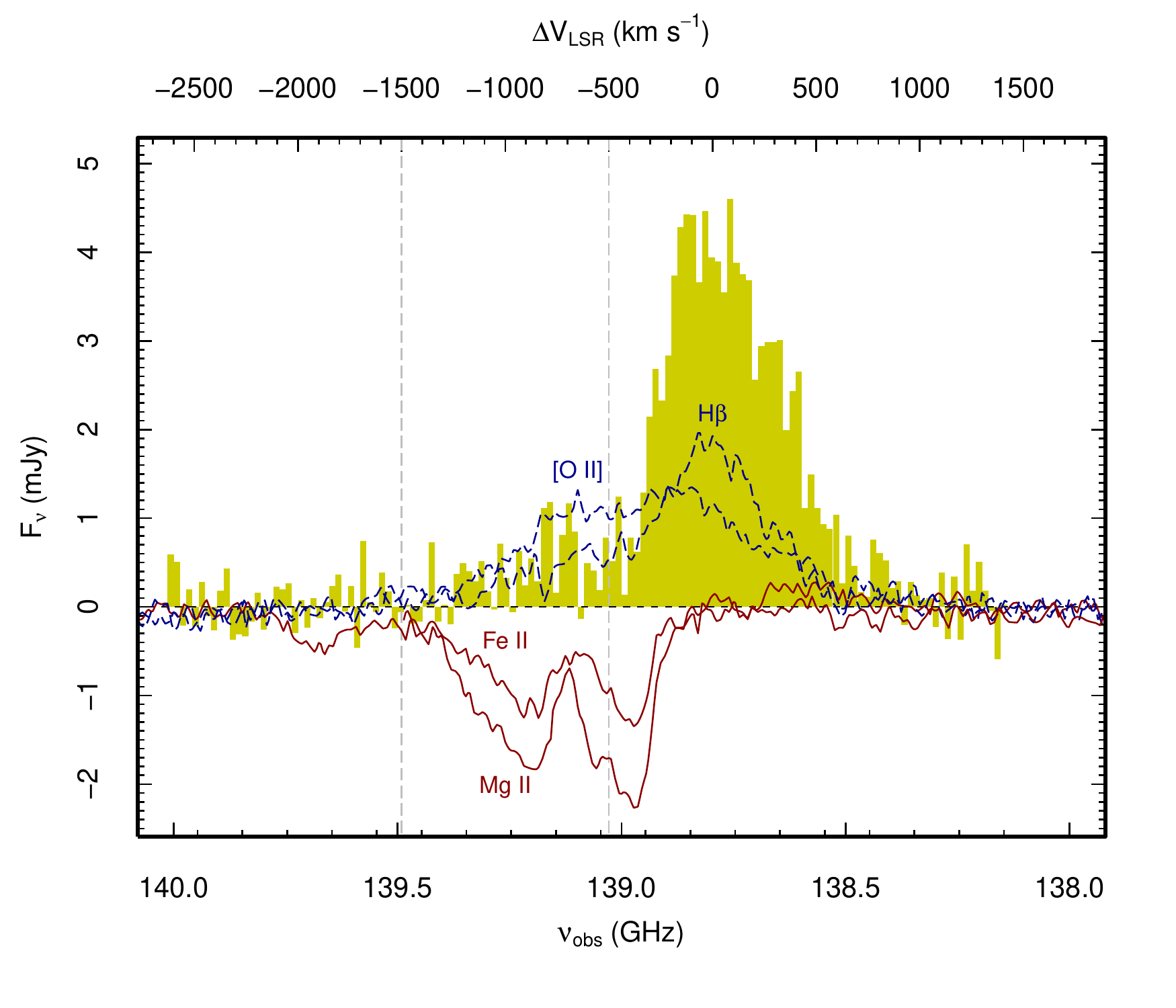}}
\caption{Total \co{} spectrum of SDSS\,J1341$-$0321, summed over the 3$\sigma$ contour of the zeroth moment image. We also
  show the continuum-subtracted MagE H$\beta$, [O~{\sc ii}], Fe\,{\sc
    ii}\,$\lambda$2600\AA{} and Mg\,{\sc ii} lines on the same velocity scale. The two lines of the Mg~{\sc ii} doublet are separated by 770\,km\,s$^{-1}$, but the maximum outflow velocity is larger than this value, causing some line profile blending. We crudely deblend the Mg~{\sc ii} doublet by showing the profile of the redder line (2804\AA{}) at $\Delta V > -770$\,km\,s$^{-1}$ and the profile of the bluer line (2796\AA{}) at $\Delta V < -770$\,km\,s$^{-1}$.  We deblend the [O~{\sc ii}] doublet in a similar manner, but the closer spacing of the two lines ($\Delta V = 222$\,km\,s$^{-1}$) means there is still some velocity overlap.  The Fe\,{\sc ii} and H$\beta$ lines do not suffer from blending.  Significant \co{} emission extends to at least $\Delta V\approx-1000$\,km\,s$^{-1}$, similar to the nebular lines. There are at least two distinct velocity components seen in the absorption lines. Vertical dashed lines indicate the velocity range we consider to be outflowing molecular gas.}
\end{figure}

\section{Conclusions and closing remarks}

Explaining the survival of cold--cool\footnote{i.e.\ molecular, atomic and ionized gas with $T<10^4$\,K} gas in fast galaxy scale winds has traditionally been a struggle for simulations, and the presence of cold molecular gas at first glance exacerbates this issue. Radiative driving was originally proposed as an attractive mechanism to propel cold gas to large galactocentric radius since it favors the survival of cold--cool clouds compared to ram pressure driving (Murray et al.\ 2005; Thompson et al. 2015), although multiple driving mechanisms are expected to be required to produce a multiphase wind. However, rather than being directly expelled, cool, and even cold, gas could condense out of a hot wind. Schneider et al.\ (2018) present hydrodynamic simulations demonstrating that, when radiative cooling is included, a hot, mass-loaded ($\eta\approx0.5$) outflow can produce cool (10$^4$\,K) gas with $v_{\rm out}\approx1000$\,km\,s$^{-1}$, but the cooling floor in these simulations precludes an analysis of any molecular component. Through an analytic approach, Zubovas \& King (2014) argue that {\it molecular} gas can cool out of a hot (AGN shock-driven) wind, and in hydro-chemical simulations Richings \& Faucher-Gigu\'ere\ (2018ab) also demonstrate that molecular outflows can form via {\it in situ} cooling within AGN-driven winds

Relevant to this work, Zubovas \& King propose that 1000\,km\,s$^{-1}$ cold--cool gas outflows {\it always} have an AGN origin even in cases -- like this -- where an AGN is not visible, since the wind signature is visible for some 10s of Myr after the central engine switches off. Given the tension between the energetics of the molecular outflow and the energy budget of the recent starburst, we leave open the possibility that SDSS\,J1341$-$0321  -- and possibly other galaxies in our sample -- presents an example of a relic AGN outflow. Nevertheless, we conclude by remarking on this galaxy's extraordinarily high implied star formation surface density of $\Sigma_{\rm SFR}\approx30,000$\,$M_\odot$\,yr$^{-1}$\,kpc$^{-2}$, and maintain that momentum injection by compact starbursts triggered during major mergers can be a competitive mechanism to quench the centres of massive galaxies. 

\section*{Acknowledgements}

We thank the anonymous referee for a constructive report that helped improve this work. We also thank Tim Heckman for useful discussions and Charlie Conroy for providing his C3K models prior to publication. J.E.G. is supported by the Royal Society. G.R. would like to thank the University of Hamburg Observatory and the European Southern Observatory (ESO) for their hospitality in hosting him while working on this paper. G.R. would additionally like to thank ESO for financial support through their scientific visitor program. The National Radio Astronomy Observatory is a facility of the National Science Foundation operated under cooperative agreement by Associated Universities, Inc. R.C.H. acknowledges support from the National Science Foundation through CAREER grant number 1554584. The authors acknowledge excellent support of the UK ALMA Regional Centre Node. This letter makes use of the following ALMA data: 2016.1.01072.S, 2017.1.0318.S. ALMA is a partnership of ESO (representing its member states), NSF (USA), and NINS (Japan), together with NRC (Canada) and NSC and ASIAA (Taiwan), in cooperation with the Republic of Chile. The Joint ALMA Observatory is operated by ESO, AUI/NRAO, and NAOJ. Support for HST-GO-12272, HST-GO-13689, and HST-GO-14239 was provided by NASA through a grant from the Space Telescope Science Institute, which is operated by the Association of Universities for Research in Astronomy, Incorporated, under NASA contract NAS5-26555.

\end{document}